\documentclass[11pt]{article}

\usepackage[preprint]{acl}

\usepackage{times}
\usepackage{latexsym}
\usepackage{enumitem}
\usepackage[T1]{fontenc}

\usepackage[utf8]{inputenc}

\usepackage{microtype}

\usepackage{inconsolata}

\usepackage{graphicx}

\usepackage[utf8]{inputenc} 
\usepackage[T1]{fontenc}    
\usepackage{hyperref}       
\usepackage{amsmath}
\usepackage[dvipsnames]{xcolor}
\usepackage[table]{xcolor}
\usepackage{url}            
\usepackage{booktabs}       
\usepackage{amsfonts}       
\usepackage{nicefrac}       
\usepackage{microtype}      
\usepackage{xcolor}         
\usepackage{multirow}
\usepackage{wrapfig}
\usepackage{array}

\definecolor{VF}{HTML}{d44232}
\definecolor{EE}{HTML}{6692b4}
\definecolor{IB}{HTML}{8eae2f}
\definecolor{TP}{HTML}{e99327}

\title{miniReranker: Efficient Multimodal Reranking through Visual Cache Reuse and Interaction Sparsity}

\author{
 \textbf{Yingqi Fan\textsuperscript{1}},
 \textbf{Xuan Lu\textsuperscript{1}},
 \textbf{Anhao Zhao\textsuperscript{1}},
 \textbf{Junlong Tong\textsuperscript{1}},
\\
 \textbf{Ping Nie\textsuperscript{2}},
 \textbf{Kai Zou\textsuperscript{2}},
 \textbf{Yunpu Ma\textsuperscript{3}},
 \textbf{Wei Zhang \textsuperscript{1}},
 \textbf{Xiaoyu Shen\textsuperscript{1}\thanks{Corresponding Author}},
\\
\\
 \textsuperscript{1}Eastern Institute of Technology, Ningbo,
 \textsuperscript{2}University of Waterloo,
\\
 \textsuperscript{3}Netmind.ai,
 \textsuperscript{4}Munich Center for Machine Learning, LMU
\\
\\
 \small{
   \textbf{Correspondence:} \href{mailto:email@domain}{yingqi949@gmail.com} \href{mailto:email@domain}{xyshen@eitech.edu.cn}
 }
}

\usepackage[most]{tcolorbox}
\usepackage{xcolor}

\definecolor{PromptFrame}{HTML}{4C78A8}
\definecolor{PromptBack}{HTML}{EAF2FB}
\definecolor{PromptTitle}{HTML}{D9E8F6}

\newtcblisting{promptbox}[1]{
    listing only,
    listing options={
        basicstyle=\ttfamily\footnotesize,
        breaklines=true,
        columns=fullflexible,
        keepspaces=true
    },
    colback=PromptBack,
    colframe=PromptFrame,
    coltitle=black,
    colbacktitle=PromptTitle,
    title=\textbf{#1},
    fonttitle=\small,
    boxrule=0.8pt,
    arc=2pt,
    left=4pt,
    right=4pt,
    top=4pt,
    bottom=4pt,
    toptitle=3pt,
    bottomtitle=3pt,
    width=\linewidth,
    enhanced
}

\begin{document}
\maketitle
\begin{abstract}
Multimodal large language models (MLLMs) have emerged as powerful rerankers for multimodal retrieval, but their fine-grained token-level interactions come at a substantial computational cost. In point-wise reranking, each query must be independently paired with many candidate documents, resulting in repeated computation over highly overlapping multimodal inputs. In this work, we propose \emph{miniReranker}, an efficient MLLM reranking framework that reduces redundancy at multiple levels. First, we introduce a \textcolor{VF}{\emph{vision-first}} reformulation that aligns with native MLLM prompting formats while maximizing reuse of expensive visual representations through prefix caching. We then identify redundancy in model depth, cross-segment interactions, and visual token representations, motivating three complementary optimizations:  \textcolor{EE}{\emph{early exit}},  \textcolor{IB}{\emph{interaction bands}}, and \textcolor{TP}{\emph{embedder-guided token pruning}}. Built on Qwen3-VL, \emph{miniReranker} achieves similar performance while reducing active parameters to 58\% and achieving nearly $3\times$ training acceleration. When reranking top-$100$ candidates, it reduces reranking runtime by up to $>$99\%, demonstrating that efficient MLLM reranking can be achieved without sacrificing quality.
\end{abstract}

\section{Introduction}

Multimodal large language models (MLLMs) have recently emerged as powerful rerankers for multimodal retrieval systems~\cite{survey:mm_rerank}. Unlike dual-encoder retrievers that independently map queries and documents into global embeddings~\cite{gme_GeneralMultimodalEmbedder}, MLLM rerankers jointly process query-document pairs through token-level interactions, enabling fine-grained cross-modal reasoning and substantially stronger relevance modeling~\cite{dai2025supervisedfinetuningcontrastivelearning,qwen3-vl-embed-rerank,liu2025lamra}. However, this stronger interaction capability comes at a significant computational cost. In modern retrieval systems, each query must be compared against dozens or hundreds of candidate documents, forcing the reranker to repeatedly process highly overlapping multimodal inputs~\cite{chen2025mllm}.

Existing point-wise MLLM rerankers predominantly inherit the query-first formulation from traditional text retrieval systems, arranging inputs as $[\mathrm{ins}, q, d]$ regardless of modality composition~\cite{Qwen3-VL,vlm-as-reranker,mm-r5,liu2025lamra,lin2025mmembed}. While natural for purely textual retrieval, this design is suboptimal for multimodal reranking from both effectiveness and efficiency perspectives. From an effectiveness standpoint, modern MLLMs are primarily pretrained on visual-question-answering (VQA) style data, where visual content naturally precedes textual instructions and questions~\cite{llava,internvl}. Query-first prompting therefore frequently breaks the model's native input format, particularly in text-to-visual retrieval settings. From an efficiency standpoint, sequence order determines which representations can be reused through prefix KV caching~\cite{qin2026prefillasaservicekvcachenextgenerationmodels}. Because visual inputs typically dominate the computational cost~\cite{lin2024preserve}, a fixed query-first formulation often forces expensive visual documents to be repeatedly encoded across candidate pairs. Simply adopting a document-first formulation is not a universal solution, as it becomes suboptimal when the query itself contains visual content. Motivated by these observations, we introduce a simple \emph{vision-first} reformulation that consistently places visual inputs before textual inputs. This strategy simultaneously restores alignment with MLLM pretraining formats and maximizes reuse of expensive visual computation, yielding improvements in both reranking effectiveness and efficiency.

Although vision-first prompting eliminates a substantial fraction of redundant computation, the remaining cost is still considerable. Beyond sequence ordering, dense MLLM rerankers perform extensive computation throughout the entire network depth, repeatedly execute cross-segment attention across all layers, and process large numbers of visual tokens whose contributions to the final relevance decision may be marginal~\cite{fan2026visualtokensreallyencode}. To better understand these inefficiencies, we analyze reranking behavior through layer-wise logit probing~\cite{earlyexit:logitlens}. Our analysis reveals that reranking signals emerge much earlier than final-layer predictions, that effective query-document interactions are concentrated within a narrow subset of layers, and that many visual tokens contribute little to the final relevance score.

Building on these observations, we propose \emph{miniReranker}, an efficient MLLM reranking framework that combines vision-first prompting with three complementary compression strategies. First, we employ \emph{early exit} to truncate unnecessary upper transformer layers after relevance signals have largely converged. Second, we introduce an \emph{interaction band} that restricts expensive query-document attention to the layers where meaningful cross-segment information exchange actually occurs. Third, we perform \emph{embedder-guided token pruning}, leveraging attention information already produced by the retrieval-stage encoder to remove redundant visual tokens without requiring additional forward passes.

We instantiate \emph{miniReranker} on top of Qwen3-VL-Instruct~\cite{Qwen3-VL} and fine-tune it using a point-wise \texttt{yes}/\texttt{no} relevance objective on a newly constructed multimodal reranking dataset. We evaluate \emph{miniReranker} on MMEB-v2~\cite{benchmark:MMEB-v2}, covering image, visual-document, and video tasks. Results show that \emph{miniReranker} outperforms the original instruct models and remains competitive with existing multimodal embedding/reranking baselines, while preserving $\sim96\%$ of the dense reranker performance.

In terms of efficiency, \emph{miniReranker} reduces active parameters to about $58\%$ of the dense model, achieves nearly $3\times$ training acceleration, and substantially lowers online reranking latency. In particular, when reranking Top-$100$ candidates with a single query, \emph{miniReranker} reduces video reranking runtime to $<1\%$ of the dense implementation and image reranking runtime to $<15\%$. Such savings become even more significant as the number of queries increases. Our main contributions are summarized as follows:

\begin{itemize}[leftmargin=12pt,itemsep=1pt,topsep=2pt,parsep=0pt]
    \item We propose a \textit{vision-first} prompt reformulation that enables reusable visual pre-caching while improving reranking effectiveness.
    \item We reveal that reranking computation in MLLMs contains substantial depth-, interaction-, and token-level redundancy.
    \item We conduct extensive experiments on MMEB-v2 across 78 tasks, showing that \texttt{miniReranker} preserves most of the dense reranking performance while reducing reranking runtime by over $99\%$ in high-reuse settings.
\end{itemize}

\section{Preliminaries}

\begin{figure*}[t]
    \centering
    \includegraphics[width=1\linewidth]{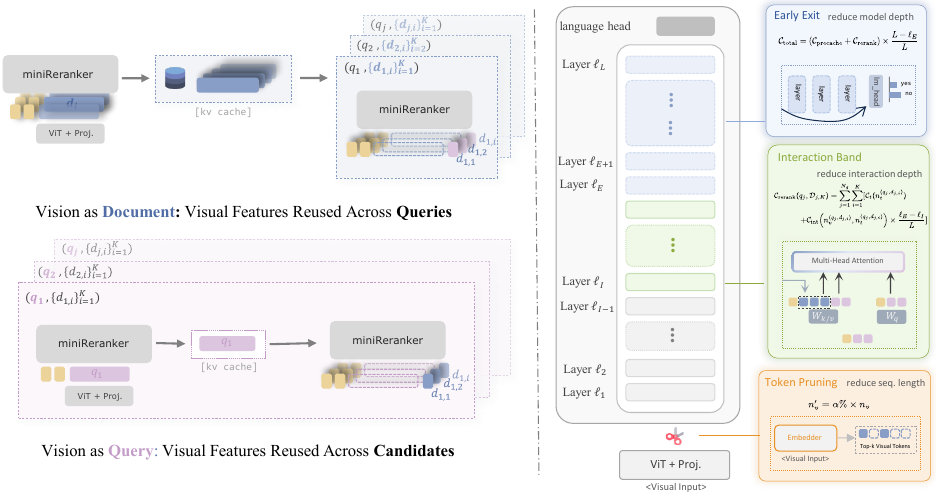}
    \vspace{-20pt}
    \caption{Overview of \textbf{miniReranker}. Left: the proposed \textcolor{VF}{\emph{Vision-first}} reformulation enables reusable visual pre-caching for both vision-as-document and vision-as-query settings. Right: \texttt{miniReranker} further improves efficiency through three complementary compression strategies: (1) \textcolor{EE}{\emph{Early Exit}}, which reduces depth-wise computation by terminating inference at intermediate layers; (2) \textcolor{IB}{\emph{Interaction Band}}, which restricts cross-segment attention to a narrow range of effective layers; and (3) \textcolor{TP}{\emph{Token Pruning}}, which removes redundant visual tokens to reduce sequence length and interaction cost.}
    \label{fig:42_compression}
    \vspace{-12pt}
\end{figure*}

Modern multimodal search systems typically follow a two-stage ``retrieve-and-rerank'' pipeline~\cite{nogueira2020passagererankingbert,nogueira2019multistagedocumentrankingbert}. Given a query $q$, a fast, coarse-grained retriever first extracts a manageable candidate set of documents $\mathcal{D}=\{d_1, \dots, d_K\}$ from a large corpus. A multimodal reranker is then tasked with precisely rescoring and reordering these candidates. Because both the query and the documents can be represented in either textual or visual (image/video) formats, this formulation accommodates diverse paradigms, including text-to-visual (T$\to$V), visual-to-text (V$\to$T), and visual-to-visual (V$\to$V)~\cite{wei2023uniirtrainingbenchmarkinguniversal}.

\subsection{Multimodal Point-wise Reranking}
While list-wise approaches evaluate multiple candidates simultaneously, they require concatenating numerous documents into a single, exceptionally long sequence~\cite{rankgpt,list_wise_rerank_1,list_wise_rerank_2}. Because modern MLLMs often suffer from severe performance degradation and attention dilution when processing long lists of interleaved visual inputs~\cite{liu-etal-2024-lost,attnsink}, the point-wise formulation has emerged as the prevailing paradigm.
\paragraph{MLLM Architecture.} To process these heterogeneous modalities, MLLMs map textual and visual inputs into a unified token space~\cite{wu2026from}. Text is tokenized directly, whereas visual inputs are encoded by a vision model (e.g., a Vision Transformer~\cite{vit}) and subsequently projected into the language model's embedding space~\cite{llava}. The pointwise MLLM reranker, parameterized by $L$ transformer layers and a hidden dimension $d_h$, then operates over a concatenated sequence comprising system instruction tokens, $n_q$ query tokens, and $n_d$ document tokens.

\paragraph{Point-wise Scoring.} 
A point-wise MLLM reranker evaluates each query--document pair entirely independently. Guided by a predefined task instruction $\mathrm{ins}$, the model estimates the relevance of a document $d$ to the query $q$ by generating a binary \texttt{yes}/\texttt{no} decision at the next-token prediction step, modeling the conditional probability $p(\cdot \mid \mathrm{ins}, q, d)$. 

Let $z_{\texttt{yes}}$ and $z_{\texttt{no}}$ denote the output logits corresponding to the \texttt{yes} and \texttt{no} tokens. The final relevance score $s(q,d)$ used to rank the candidates is computed as the normalized probability of the positive class~\cite{qwen3-vl-embed-rerank,dai2025supervisedfinetuningcontrastivelearning}:
\begin{equation}
s(q,d)=
\sigma(z_{\texttt{yes}}-z_{\texttt{no}}),
\label{eq:rerank}
\end{equation}
where $\sigma(\cdot)$ is the sigmoid function.

\subsection{Complexity Analysis}
\label{subsection:complexity_analysis}
The sequence order of the query and document dictates how much of this computation must be performed online. Because the task instruction $\mathrm{ins}$ is fixed across queries, its representations can always be pre-cached offline; thus, we omit it from the online complexity analysis.

\paragraph{Query-First Configuration.} MLLM rerankers can adopt a query-first sequence order, formatted as $[\mathrm{ins}, q, d]$. Because the query $q$ is identical for all $K$ candidates for a given user request, its key-value representations and FFN outputs can be computed once and broadcasted. However, because $q$ precedes $d$, the document tokens $d$ are dynamic relative to the prefix and must be encoded online $K$ times. For each candidate, the $n_d$ document tokens attend to the query prefix and themselves, and pass through the FFN. The total online complexity to rerank $K$ documents is therefore:
\begin{equation*}
\begin{split}
\mathcal{C}_{\mathrm{q-first}} = \mathcal{O}\Big( & L \big[n_q^2 d_{h} + n_q d_{h}^2\big] \\
& + K L \big[n_d (n_q + n_d) d_{h} + n_d d_{h}^2\big]\Big).
\end{split}
\end{equation*}
The first line represents the one-time query encoding, while the second line captures the repeated self/cross-attention and FFN operations of the document tokens.

\paragraph{Document-First Configuration.} Alternatively, a document-first configuration orders the input as $[\mathrm{ins}, d, q]$. Because the candidate documents are often fixed in a corpus, the representations of the document tokens $d$ (both their self-attention keys/values and their FFN transformations) can be pre-computed and cached offline. During online inference, for each of the $K$ candidates, the MLLM only needs to encode the $n_q$ query tokens. These query tokens perform attention against the cached document representations and pass through the FFN to generate the \texttt{yes}/\texttt{no} logits. This reduces the online computational complexity to:
\begin{equation*}
\mathcal{C}_{\mathrm{d-first}} = \mathcal{O}\Big(K L \big[n_q (n_d + n_q) d_{h} + n_q d_{h}^2\big]\Big).
\end{equation*}
\section{Methods}

We propose an efficient framework that optimizes MLLM reranking through two modifications. First, a \textit{vision-first} reformulation (Sec.~\ref{subsec:vision-first}) aligns the input with native MLLM pre-training formats while maximizing the reuse of heavy visual computations. Second, to reduce redundancy in cross-modal information flows, we introduce three inference-time optimizations: early exiting, reduced interaction bands, and token pruning (Sec.~\ref{subsec:compression}).

\definecolor{deepviolet}{HTML}{3D2A7A}
\definecolor{darkgreen}{HTML}{26614f}
\begin{figure*}[t]
    \centering
    \includegraphics[width=1\linewidth]{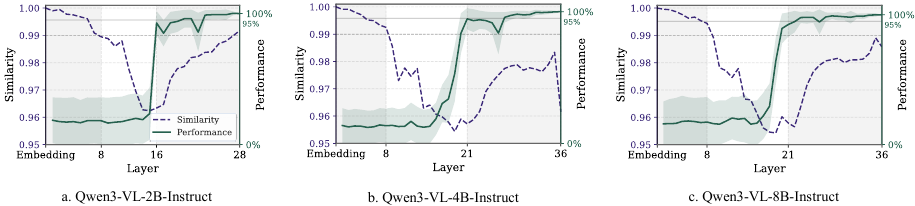}
    \vspace{-10pt}
    \caption{\textbf{\textcolor{deepviolet}{Layer-wise Logit Probing}} reveals substantial depth-wise redundancy in multimodal reranking, while \textbf{\textcolor{darkgreen}{Cross-segment Interaction Analysis}} shows that effective cross-segment information exchange is concentrated within a narrow range of intermediate layers.}
    \label{fig:42_performance_similarity}
    \vspace{-12pt}
\end{figure*}

\subsection{Vision-First Prompt Reformulation}
\label{subsec:vision-first}
\paragraph{Limitations of Query-First Rerankers.} 
Existing point-wise MLLM rerankers predominantly inherit the \textit{query-first} configuration from traditional text-based retrieval~\cite{Qwen3-VL,vlm-as-reranker,mm-r5,liu2025lamra,lin2025mmembed,qwen2vl}, strictly formatting the input sequence as $[\mathrm{ins}, q, d]$. However, blindly applying this rigid, text-centric design to multimodal models introduces two major bottlenecks. First, from an accuracy perspective, modern MLLMs are typically pre-trained on Visual Question Answering (VQA) formats~\cite{llava,instructblip,flamingo,internvl}, where the model is optimized to process visual tokens \textit{before} textual ones. In the common T$\to$V setting (text query, visual document), a query-first setup forces the MLLM to process the text query before the visual document, directly breaking its native pre-training alignment. Second, from an efficiency perspective, sequence order dictates which segment must be recomputed $K$ times online (Sec.~\ref{subsection:complexity_analysis}). Because visual tokens often dominate the computational cost, the query-first setup forces heavy visual documents in the T$\to$V setting to be repeatedly re-encoded online for every candidate. Conversely, rigidly adopting a document-first setup to fix this would create a massive bottleneck in the V$\to$T setting, where the heavy visual query would then have to be repeatedly encoded.

\paragraph{Dynamic Sequence Ordering.}
To resolve both the accuracy and efficiency bottlenecks simultaneously, we advocate for a dynamic configuration that strictly places the \emph{visual modality ahead of the textual modality}. Concretely, for the T$\to$V setting, we adopt a \textit{document-first} order $[\mathrm{ins}, d, q]$, which restores the natural VQA prompting style for higher accuracy and allows the heavy visual documents to be cached offline. For the V$\to$T setting, we adopt a \textit{query-first} order $[\mathrm{ins}, q, d]$, which maintains the VQA alignment (vision precedes text) and ensures the massive visual query is encoded only once per user request and broadcasted. In the V$\to$V setting where both inputs are visual, we place the document first. We empirically find it performs similarly to query-first in terms of accuracy, but offers the critical system advantage of enabling offline pre-caching for the static document corpus. As shown in Tab.~\ref{tab1:rerank_format}, adhering to this format consistently yields superior performance both in zero-shot settings (Instruct) and after supervised fine-tuning (reranker$^{\dagger}$)\footnote{${\dagger}$ denotes models fine-tuned using our datasets; experimental details are provided in Sec.~\ref{sec:experiments}.}.

\begin{table}[h]
\vspace{-5pt}
\small
\centering
\begin{tabular}{lcccc}
\toprule
\multirow{2}{*}{Model}  & \multicolumn{2}{c}{2B}  & 4B & 8B \\
\cmidrule(lr){2-3} \cmidrule(lr){4-4} \cmidrule(lr){5-5}
 & Instruct & reranker$^{\dagger}$ & Instruct & Instruct \\
\midrule
Q-first  & 61.5 & 71.3 & 67.5 & 69.7  \\ 
D-first  & 52.7 & 69.6 & 61.9 & 64.2  \\
\rowcolor{VF!15} V-first  & \textbf{63.6} & \textbf{73.0} & \textbf{68.8} & \textbf{70.8}  \\
\bottomrule
\end{tabular}
\vspace{-5pt}
\caption{Prompt format comparison.}
\vspace{-12pt}
\label{tab1:rerank_format}
\end{table}

\paragraph{Theoretical FLOPs Reduction.}
We quantify the exact computational savings ($\Delta \mathcal{C}$) of our dynamic formulation by evaluating the difference between the optimal and sub-optimal sequence orders. By expanding the equations from Sec.~\ref{subsection:complexity_analysis}, the cross-attention interaction cost ($\mathcal{O}(K L n_q n_d d_h)$) perfectly cancels out during subtraction, beautifully isolating the pure FLOPs saved in self-attention and FFN encoding. In the \textbf{T$\to$V setting} ($n_q \ll n_d$), comparing our vision-first (document-first) approach against the standard query-first baseline yields an exact online reduction of $\Delta \mathcal{C}_{\mathrm{T\to V}} = \mathcal{C}_{\mathrm{q-first}} - \mathcal{C}_{\mathrm{d-first}} = \mathcal{O}\big( L d_h [ K n_d(n_d+d_h) - (K-1) n_q(n_q+d_h) ] \big)$. Because the visual document is massive ($n_d \gg n_q$), the positive term overwhelmingly dominates, demonstrating the elimination of $K$ repeated online encodings of the heavy visual document. Conversely, in the \textbf{V$\to$T setting} ($n_q \gg n_d$), our vision-first formulation equals a query-first setup. Compared to a rigid document-first strategy, the reduction in online FLOPs is $\Delta \mathcal{C}_{\mathrm{V\to T}} = \mathcal{C}_{\mathrm{d-first}} - \mathcal{C}_{\mathrm{q-first}} = \mathcal{O}\big( L d_h [ (K-1) n_q(n_q+d_h) - K n_d(n_d+d_h) ] \big)$. By broadcasting the visual query, we successfully avoid $K-1$ redundant online encodings of the massive visual tokens. Finally, in the \textbf{V$\to$V setting} where both modalities are heavy, our vision-first (document-first) approach exploits global corpus redundancy. Across a system serving $N_q$ queries over a static corpus of $N_d$ documents, the query-first baseline re-encodes visual documents $N_q K$ times. Pre-caching these documents reduces the system-level document encoding FLOPs by exactly $\Delta \mathcal{C}_{\mathrm{V\to V}} = \mathcal{O}\big( (N_q K - N_d) L n_d(n_d+d_h)d_h \big)$, yielding immense computational relief at scale.

\subsection{Model-Level Compression}
\label{subsec:compression}

While the vision-first prompt reformulation optimizes cross-modal sequence ordering, MLLMs still suffer from massive internal computational overhead during the forward pass. To systematically eliminate this redundancy, we analyze the reranking process from a model-level perspective and introduce three orthogonal compression strategies targeting model depth, cross-segment attention density, and sequence length.

\paragraph{\textcolor{EE}{Early Exit}: Truncating Model Depth.}
Unlike open-ended visual generation, which depends on deep iterative reasoning to autoregressively decode precise output tokens, point-wise reranking only requires the model to estimate a relevance score from the prefill representations. Prior studies have shown that the deepest layers of large language models are mainly responsible for linguistic refinement and final token disambiguation, transforming internal representations into fluent natural language outputs~\cite{earlyexit:dola,earlyexit:visipruner,earlyexit:stagesofllm,ealyexit:layerbylayer,liu2026vicaefficientmultimodalllms}. Since reranking does not require autoregressive text generation, these late-stage decoding capabilities might be largely unnecessary for the task.

To validate this assumption, we conduct a layer-wise probing analysis using the logit lens technique~\cite{earlyexit:logitlens}. At each transformer layer $l$, we extract the hidden state $h^{(l)}$ at the final token position, apply the final layer normalization $\mathrm{LN}$, and project it to the vocabulary space $\mathcal{X}$ using the model's output head $\phi(\cdot)$:
\begin{equation}
\vspace{-3pt}
z^{(l)}_x = \phi(\mathrm{LN}(h^{(l)}))_x, \quad x \in \mathcal{X}.
\end{equation}
We then compute the relevance score using the \texttt{yes}/\texttt{no} logits via Eq.~\ref{eq:rerank}. As shown in Fig.~\ref{fig:42_performance_similarity}, reranking accuracy rises sharply in the intermediate layers and saturates early, achieving $\sim$95\% of the final-layer performance using only $<$60\% of the network depth.~\footnote{See App.~\ref{appsubsec:early_exit_vqa} for comparisons with general VQA tasks.} Motivated by this, we employ an early-exit strategy that safely terminates the forward pass at a predefined intermediate layer, significantly reducing depth-wise computation without meaningful accuracy loss.

\paragraph{\textcolor{IB}{Interaction Band}: Localizing Query-Document Attention.}
Fig.~\ref{fig:42_performance_similarity} reveals that scoring capabilities emerge abruptly within a narrow subset of layers, implying that effective query-document interaction might also be highly localized. To reduce computational redundancy, we first investigate how relevance signals from the first segment (query or document, depending on sequence order) reach the final decision token. By selectively masking attention, we compare a \emph{direct pathway} (final token attends directly to the first segment) against an \emph{indirect pathway} (information fuses into the second segment, which then informs the final token). Blocking the direct pathway barely affects accuracy (61.5$\rightarrow$61.0), whereas blocking the indirect pathway (disabling query-document attention) causes a catastrophic collapse (61.5$\rightarrow$5.5). This confirms that relevance information must be fused into the second segment before reaching the final token.

To identify \emph{where} this critical fusion occurs, we fix the second segment, vary the first, and measure the cosine similarity of the second segment's internal representations across layers (Fig.~\ref{fig:42_performance_similarity}). In shallow layers, similarity remains near 1.0, indicating independent evolution. As depth progresses, similarity drops sharply, marking the exact onset of active cross-modal integration. Crucially, this divergence perfectly aligns with the accuracy spike observed in our early-exit analysis. We define this concentrated region of effective query-document attention as the \emph{interaction band}.

To exploit this localized integration, we design a selective interaction mechanism. During training, a sparse attention mask forces segments to evolve independently in non-interaction layers by blocking query-document attention, retaining standard causal attention solely within the interaction band. Consequently, during inference, we only load the first segment's key-value (KV) cache for this specific subset of layers. This preserves essential cross-segment information exchange while safely eliminating expensive quadratic attention operations elsewhere. Empirically, we set this interaction band to layers 8--16 for Qwen3-VL-2B, and layers 8--21 for Qwen3-VL-4B/8B.

\paragraph{\textcolor{TP}{Token Pruning}: Condensing Visual Sequences.}
While early exiting and the interaction band effectively reduce computational depth and cross-attention density, the raw sequence length---typically dominated by massive numbers of visual patches---remains a fundamental bottleneck. To directly compress the visual sequence before it enters the MLLM backbone, we introduce an embedder-guided token pruning strategy.

Rather than introducing expensive operations to dynamically evaluate token importance during the reranking pass, we obtain these importance scores entirely for free. Because the visual inputs have already been processed by an upstream first-stage retriever (e.g., \texttt{Qwen3-VL-Embedding}) during the initial recall phase, we simply repurpose its readily available computation. Specifically, we extract the attention weights directed from the embedder's final sequence token to the visual tokens. However, because attention patterns fluctuate drastically across different network depths~\cite{attn_pruning:VisAttnSinks,attn_pruning:PACT,attn_pruning:SGL,attn_pruning:VISA,wu2026hidrop}, relying on a single layer (e.g., the 3rd layer~\cite{fastv}) yields unstable token selection. To ensure robust semantic representation, we aggregate the attention scores across \emph{all} layers of the first-stage embedder.~\footnote{See App.~\ref{appsec:selector} for additional token selection baselines.}

\begin{table}[h]
\footnotesize
\centering
\vspace{-5pt}
\begin{tabular}{lcccc}
\midrule
Selection & \texttt{Image} & \texttt{VisDoc} & \texttt{Video} & overall \\
\midrule
\rowcolor{gray!15} Dense & 59.1 & 74.4 & 49.4 & 61.5  \\
3rd Layer  & 53.2 & 70.2 & 47.9 & 57.2  \\
14th Layer & 57.3 & 71.6 & 49.0 & 59.7  \\
Last Layer & 56.8 & 72.4 & \textbf{49.5} & 59.9  \\
\rowcolor{TP!15} \textbf{All Layers} & \textbf{58.2} & \textbf{73.0} & 49.2 & \textbf{60.7}  \\
\bottomrule
\end{tabular}
\vspace{-4pt}
\caption{Comparison of visual token selection strategies under the same pruning ratio.}
\label{Tab:embedder_selection}
\end{table}

Based on these aggregated scores, we select the top $\alpha\%$ visual tokens and discard the remainder before passing the sequence to the reranker. As shown in Tab.~\ref{Tab:embedder_selection}, compared to single-layer heuristics, this layer-aggregated approach accurately preserves the most salient semantic matching features. It integrates seamlessly into the standard retrieval pipeline with zero computational overhead, allowing us to achieve an aggressive 50\% visual token compression rate while successfully recovering near-dense baseline accuracy.
\section{Experiments}
\label{sec:experiments}

\definecolor{myred}{HTML}{c83e30}
\definecolor{mygreen}{HTML}{5a9d42}

\begin{table*}[t]

\centering
\renewcommand{\arraystretch}{1.2}
\resizebox{\textwidth}{!}{
\begin{tabular}{l ccccc ccccc ccccc c}
\toprule
\multirow{2}{*}{\textbf{Model}} 
& \multicolumn{5}{c}{\textbf{Image}} 
& \multicolumn{5}{c}{\textbf{Video}} 
& \multicolumn{5}{c}{\textbf{VisDoc}}
& \multirow{2}{*}{\textbf{All}}\\
\cmidrule(lr){2-6} \cmidrule(lr){7-11} \cmidrule(lr){12-16}
& \textbf{CLS} & \textbf{QA} & \textbf{RET} & \textbf{GD} & \textbf{Overall} 
& \textbf{CLS} & \textbf{QA} & \textbf{RET} & \textbf{MRET} & \textbf{Overall} 
& \textbf{VDRv1} & \textbf{VDRv2} & \textbf{VR} & \textbf{OOD} & \textbf{Overall} \\
\midrule
ColPali v1.3 (3B)~\cite{dataset:vidore} & 
40.3 & 11.5 & 48.1 & 40.3 & 34.9 & 
26.7 & 37.8 & 21.6 & 25.5 & 28.2 & 
83.6 & 52.0 & 81.1 & 43.1 & 71.0 & 
44.4 \\
GME (2B)~\cite{gme_GeneralMultimodalEmbedder} & 
54.4 & 29.9 & 66.9 & 55.5 & 51.9 & 
34.9 & 42.0 & 25.6 & 32.4 & 33.9 & 
86.1 & 54.0 & 82.5 & 43.1 & 72.7 & 
54.1      \\
VLM2Vec (2B)~\cite{benchmark:MMEB-v2}      & 
58.7 & 49.3 & 65.0 & 72.9 & 59.7 & 
33.4 & 30.5 & 20.6 & 33.0 & 29.0 & 
49.8 & 13.5 & 51.8 & 33.5 & 41.6 & 
47.0    \\
Qwen3-VL-Reranker-2B~\cite{qwen3-vl-embed-rerank} &
---- & ---- & ---- & ---- & 74.0 & 
---- & ---- & ---- & ---- & 53.2 & 
---- & ---- & ---- & ---- & 83.2 & 
75.2 \\
Qwen3-VL-2B-Instruct~\cite{Qwen3-VL} & 
59.2 & 61.2 & 60.1 & 74.2 & 61.7 & 
60.2 & 55.4 & 20.9 & 36.7 & 44.0 & 
87.2 & 71.1 & 83.8 & 72.4 & 81.2 & 
63.6
\\
\rowcolor{VF!15} Qwen3-VL-2B-reranker$^\dagger$   & 
67.1 & 71.8 & 71.8 & 82.2 & 71.6 & 
63.8 & 60.3 & 57.8 & 49.5 & 58.8 &
93.6 & 66.7 & 92.1 & 75.5 & 85.7 & 
73.0
\\
\rowcolor{VF!15} Qwen3-VL-2B-miniReranker$^\dagger$ & 
65.6 & 65.7 & 69.7 & 76.5 & 68.2 & 
62.5 & 50.8 & 57.6 & 45.2 & 55.0 & 
93.3 & 64.3 & 92.1 & 73.8 & 84.9 & 
70.3${\textcolor{mygreen}{(96.3\%)}}$
\\
\midrule
Qwen3-VL-4B-reranker$^\dagger$   & 
69.7 & 76.1 & 75.2 & 88.7 & 75.4 & 
63.2 & 63.6 & 60.6 & 48.3 & 60.1 & 
94.3 & 66.8 & 93.9 & 75.9 & 86.5 & 
75.3
\\
Qwen3-VL-4B-miniReranker$^\dagger$   & 
68.3 & 71.5 & 73.5 & 81.8 & 72.5 &  
64.5 & 64.0 & 59.2 & 52.7 & 60.9 & 
93.8 & 68.1 & 92.7 & 75.3 & 86.2 & 
74.0${\textcolor{mygreen}{(98.3\%)}}$
\\
\midrule
GME (7B)          & 
57.7 & 34.7 & 71.2 & 59.3 & 56.0 & 
37.4 & 50.4 & 28.4 & 38.2 & 38.6 & 
89.4 & 55.6 & 85.0 & 44.4 & 75.2 & 
57.8      \\
LamRA-Qwen2.5 (7B)        & 
51.7 & 34.1 & 66.9 & 56.7 & 52.4 & 
32.9 & 42.6 & 23.2 & 37.6 & 33.7 & 
56.3 & 33.3 & 58.2 & 40.1 & 50.2 & 
47.4 \\
VLM2Vec-Qwen2VL (7B)      & 
62.7 & 56.9 & 69.4 & 82.2 & 65.5 & 
39.1 & 30.0 & 29.0 & 40.6 & 34.0 & 
56.9 & 9.4 & 59.1 & 38.1 & 46.4 & 
52.3 \\
Qwen3-VL-Reranker-8B &
---- & ---- & ---- & ---- & 78.2 & 
---- & ---- & ---- & ---- & 61.0 & 
---- & ---- & ---- & ---- & 85.8 & 
79.2 \\
\rowcolor{VF!15} Qwen3-VL-8B-reranker$^\dagger$   & 
70.0 & 77.8 & 75.2 & 90.8 & 76.2 & 
65.6 & 67.1 & 61.7 & 53.3 & 62.9 & 
94.8 & 68.2 & 94.0 & 76.4 & 87.1 & 
76.5
\\
\rowcolor{VF!15} Qwen3-VL-8B-miniReranker$^\dagger$   & 
69.2 & 73.9 & 73.7 & 89.7 & 74.3 &  
67.2 & 67.3 & 60.9 & 48.0 & 62.5 & 
94.1 & 68.6 & 93.4 & 76.3 & 86.7 & 
75.4${\textcolor{mygreen}{(98.6\%)}}$
\\

\bottomrule
\end{tabular}
}
\vspace{-5pt}
\caption{\small \textbf{Main results} on MMEB-v2 multimodal reranking benchmarks across image, video, and visual-document tasks. CLS: classification, QA: question answering. RET: retrieval, GD: grounding, MRET: moment retrieval, VDR: ViDoRe, VR: VisRAG, OOD: out-of-domain. $\dagger$ denotes models finetuned using our training recipe.}

\label{tab:main_exp}
\end{table*}

\begin{table*}[t]

\vspace{-5pt}
\centering
\renewcommand{\arraystretch}{1.2}
\resizebox{\textwidth}{!}{
\begin{tabular}{l ccccc ccccc ccccc c}
\toprule
\multirow{2}{*}{\textbf{Model}} 
& \multicolumn{5}{c}{\textbf{Image}} 
& \multicolumn{5}{c}{\textbf{Video}} 
& \multicolumn{5}{c}{\textbf{VisDoc}}
& \multirow{2}{*}{\textbf{All}}\\
\cmidrule(lr){2-6} \cmidrule(lr){7-11} \cmidrule(lr){12-16}
& \textbf{CLS} & \textbf{QA} & \textbf{RET} & \textbf{GD} & \textbf{Overall} 
& \textbf{CLS} & \textbf{QA} & \textbf{RET} & \textbf{MRET} & \textbf{Overall} 
& \textbf{VDRv1} & \textbf{VDRv2} & \textbf{VR} & \textbf{OOD} & \textbf{Overall} \\
\midrule
\rowcolor{gray!15} \multicolumn{17}{c}{\textbf{Prompting Format}}
\\
Qwen3-VL-2B-reranker$^{VF}$   & 
67.1 & 71.8 & 71.8 & 82.2 & \textbf{71.6} & 
63.8 & 60.3 & 57.8 & 49.5 & \textbf{58.8} &
93.6 & 66.7 & 92.1 & 75.5 & \textbf{85.7} & 
\textbf{73.0}
\\
Qwen3-VL-2B-reranker$^{QF}$   & 
67.1 & 70.9 & 70.5 & 84.1 & 71.2 & 
63.1 & 58.4 & 47.7 & 39.4 & 53.6 & 
91.7 & 67.5 & 91.5 & 73.6 & 84.6 & 
71.3$_{\textcolor{myred}{\downarrow 1.7}}$
\\
Qwen3-VL-2B-reranker$^{DF}$   & 
66.8 & 62.2 & 71.5 & 84.9 & 69.1 & 
59.1 & 31.8 & 57.0 & 50.9 & 49.5 & 
93.4 & 65.4 & 92.3 & 75.5 & 85.5 & 
69.6$_{\textcolor{myred}{\downarrow 3.4}}$
\\
\midrule
\rowcolor{gray!15} \multicolumn{17}{c}{\textbf{Compression Components}}
\\
Qwen3-VL-2B-reranker$^{EE}$   & 
67.0 & 69.7 & 72.0 & 83.1 & 71.3 & 
61.8 & 56.7 & 56.6 & 47.9 & 56.6 & 
93.6 & 67.1 & 92.3 & 75.3 & 85.8 & 
72.3$_{\textcolor{mygreen}{(99.0\%)}}$
\\
Qwen3-VL-2B-reranker$^{TP}$   & 
66.9 & 71.6 & 71.3 & 79.3 & 71.0 & 
64.7 & 59.6 & 57.5 & 50.4 & 58.9 & 
93.5 & 66.1 & 92.1 & 74.5 & 85.4 & 
72.7$_{\textcolor{mygreen}{(99.6\%)}}$
\\
Qwen3-VL-2B-reranker$^{IB}$   & 
67.2 & 69.9 & 71.6 & 77.3 & 70.5 & 
62.1 & 53.9 & 59.0 & 47.4 & 56.5 & 
93.7 & 63.9 & 92.4 & 74.5 & 85.2 & 
71.8$_{\textcolor{mygreen}{(98.4\%)}}$
\\

\bottomrule
\end{tabular}

}
\vspace{-5pt}
\caption{\small \textbf{Ablation study} of prompting formulations and individual compression components after supervised fine-tuning.}
\label{tab:ablation}
\vspace{-10pt}
\end{table*}

\subsection{Experimental Setup}

\paragraph{Training.} We construct a point-wise reranking training dataset covering three major multimodal categories: (i) image, (ii) visual document, and (iii) video, resulting in a total of approximately 736K training pairs. We fine-tune only the LLM component of \texttt{Qwen3-VL-Instruct} for one epoch with a supervised point-wise reranking objective, using LoRA adaptation and a learning rate of $1\times10^{-4}$; more details are provided in App.~\ref{appsec:train}. 

\paragraph{Compression Setting.}
We configure \texttt{miniReranker} according to the empirical analysis in the previous section. Specifically, the early-exit layer is set to Layer 16 for the 2B model and Layer 21 for the 4B/8B models. Across all model sizes, cross-segment interaction is enabled from Layer 8 onward, and only half of visual tokens are retained after pruning.

\paragraph{Evaluation.} 
We evaluate on MMEB-v2, covering 36 image tasks, 24 visual document tasks, and 18 video tasks~\cite{benchmark:MMEB-v2}. Reranking is performed over 1 positive sample and 19 hard negatives retrieved by \texttt{Qwen3-VL-Embedding-2B}. For image and video tasks, we report Hit@1, while for visual document tasks, we report NDCG@5. More details are provided in App.~\ref{appsec:eval}.

\subsection{Main Results}

\paragraph{Validating fine-tuning effectiveness.}
As shown in Tab.~\ref{tab:main_exp}, our finetuned point-wise rerankers consistently outperform their corresponding instruct models and achieve strong overall performance compared with existing multimodal embedding and reranking baselines at similar scales, establishing a strong dense reranking baseline for subsequent compression experiments.

\paragraph{\texttt{miniReranker}.}
Compared with the corresponding dense reranker trained under the same setup, \texttt{miniReranker} preserves $>96\%$ of the reranking performance for the 2B model and $>98\%$ for the 4B/8B models, while substantially reducing computation. The proposed compression strategy shows consistent effectiveness across model scales.

\subsection{Ablation Study}
Tab.~\ref{tab:ablation} reports ablations of prompting formulations and compression components after training.

\paragraph{Prompting Formulation.}
Among query-first (QF), document-first (DF), and the proposed vision-first (VF) prompting, VF achieves the best overall reranking performance, confirming that the vision-first formulation remains effective even after reranking finetuning.

\paragraph{Compression Components.}
We further evaluate each compression strategy independently based on the vision-first formulation. Early exit (EE), interaction band restriction (IB), and embedder-guided token pruning (TP) all preserve competitive reranking performance, showing that each component is individually effective for improving efficiency.

\section{Efficiency Analysis}
\label{sec:eff_ana}

We analyze the efficiency of 2B-miniReranker from three perspectives: parameter usage, training efficiency, and reranking latency.

\paragraph{Parameter Usage.} Our early exit strategy directly reduces the number of active parameters during both training and inference. For the 2B model, only the first 16 out of 28 layers are used, corresponding to $57.1\%$ of the original parameters (1.14B active parameters). For the 4B and 8B models, only 21 out of 36 layers are retained, resulting in $58.3\%$ parameter usage (2.2B active parameters for the 4B model and 4.6B for the 8B model). 

\begin{wrapfigure}{r}{0.5\linewidth}
    \hspace{-30pt}
    \centering
    \includegraphics[width=\linewidth]{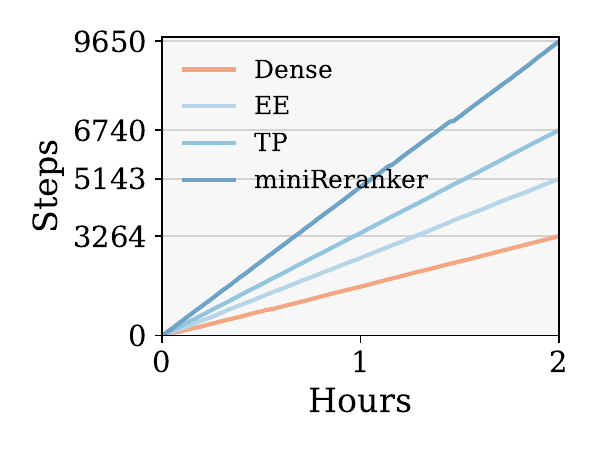}
    \vspace{-10pt}
    \caption{Training throughput comparison.}
    \vspace{-15pt}
    \label{fig:60_training_speed_comparison}
\end{wrapfigure}

\paragraph{Training Hours.} Our compression framework mitigates multimodal reranker training bottlenecks via: (1) \emph{early exit}, reduce the number of updated parameters; and (2) \emph{visual token pruning}, shorten the long multimodal sequences. These optimizations jointly reduce both forward and backward computation costs, \emph{miniReranker} achieves nearly $3\times$ training acceleration compared with the dense reranker.

\begin{figure}[h]
    \centering
    \includegraphics[width=1\linewidth]{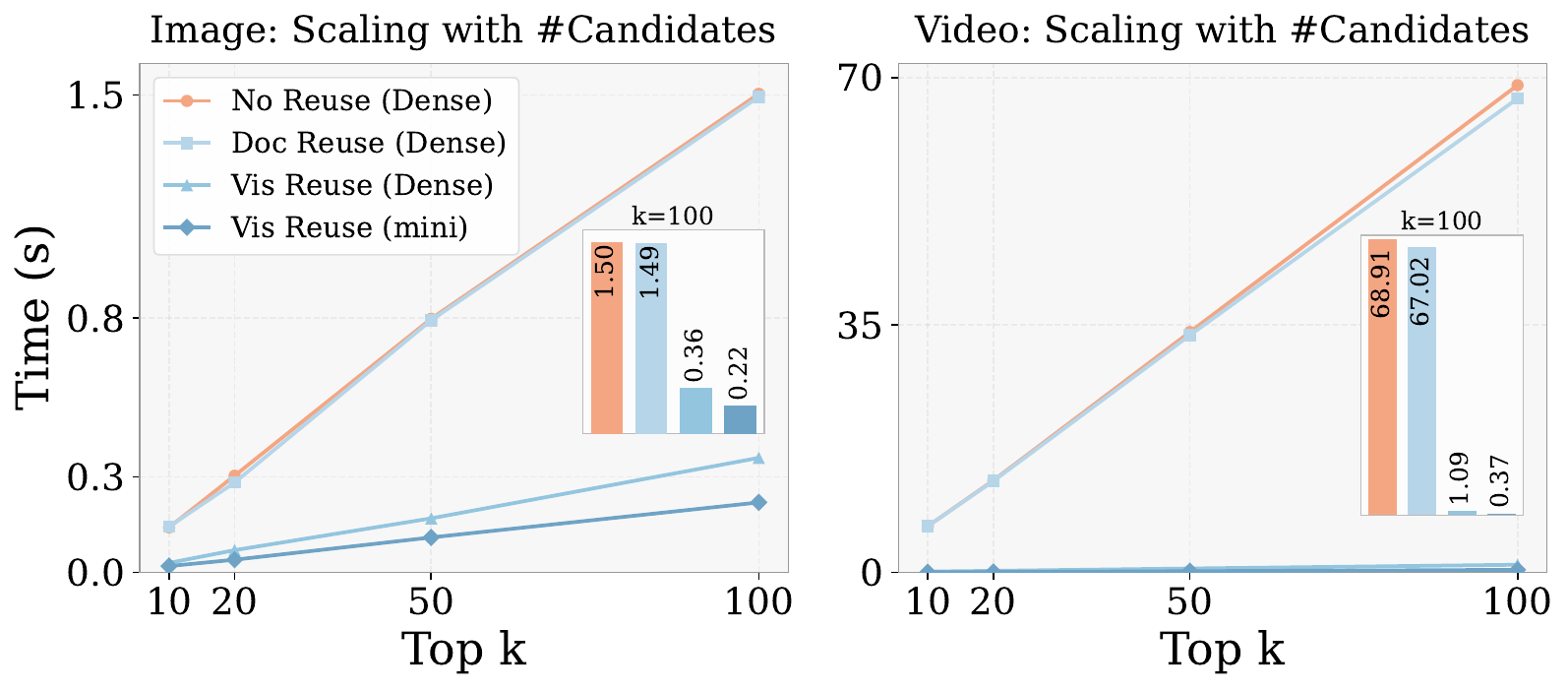}
    \vspace{-20pt}
    \caption{\small Latency scaling in the \emph{vision-as-query} setting.}
    \vspace{-10pt}
    \label{fig:61_vision_as_query}
\end{figure}

\paragraph{Reranking Latency: Vision as Query.}
For the vision-as-query setting, we evaluate on image task MS~COCO$_{\text{i2t}}$ and the video task UCF101. We fix the number of queries to 1 and gradually increase Top-$k$ from 10 to 100. As shown in Fig.~\ref{fig:61_vision_as_query}, miniReranker consistently achieves substantially lower latency than the vanilla implementation and the dense visual-reuse baseline. For MS~COCO$_{\text{i2t}}$, reranking Top-100 candidates requires only around $15\%$ of the original latency. For UCF101, which involves significantly longer video sequences, the latency is further reduced to $<1\%$ of the original runtime. Moreover, compared with the dense visual-reuse baseline, the compressed \texttt{miniReranker} further reduces latency by $\sim$66\%.

\begin{figure}[h]
    \centering
    \includegraphics[width=1\linewidth]{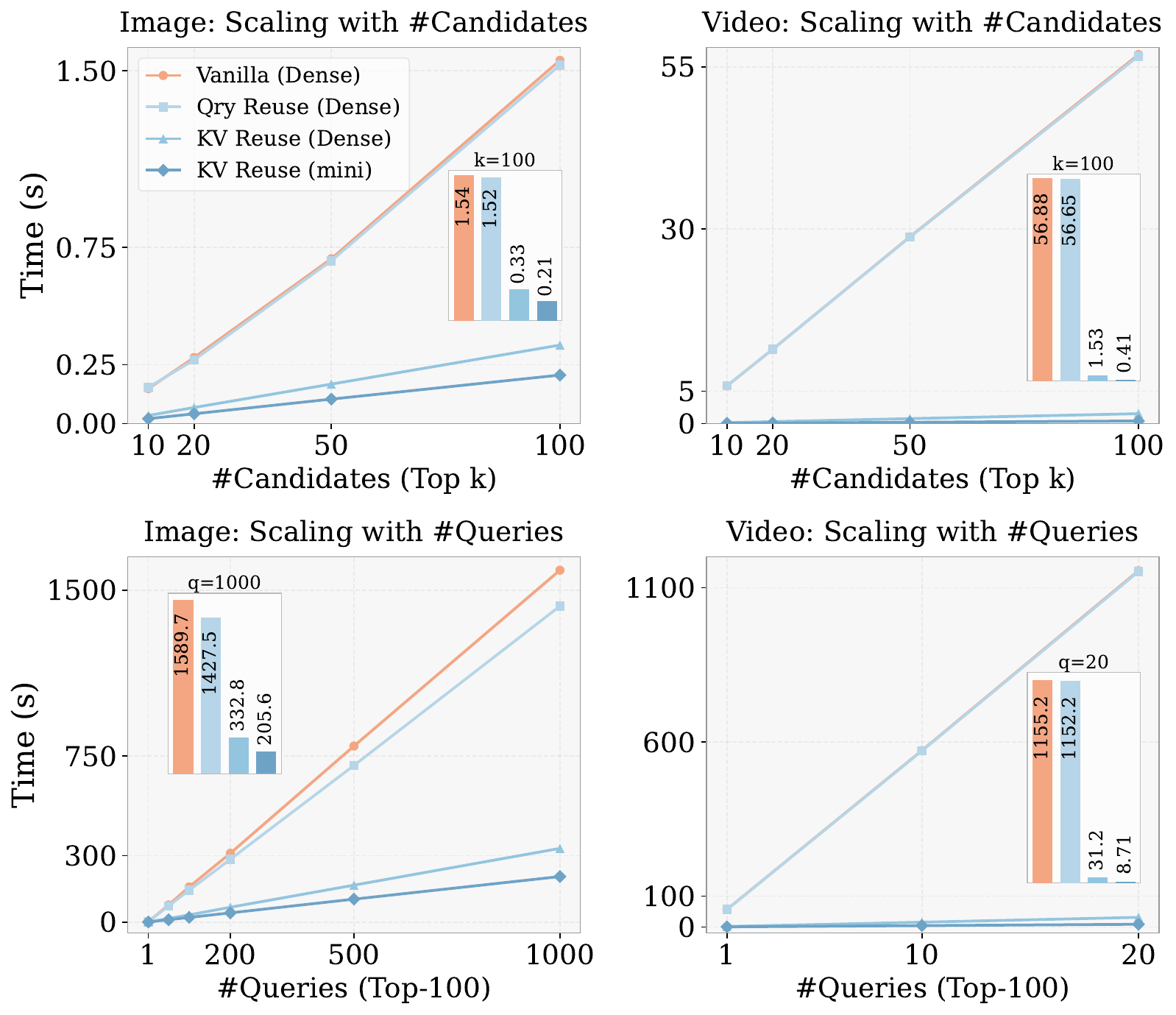}
    \vspace{-20pt}
    \caption{\small Latency scaling in the \emph{vision-as-document} setting.}
    \vspace{-10pt}
    \label{fig:62_vision_as_document}
\end{figure}

\paragraph{Reranking Latency: Vision as Document.}
For the vision-as-document setting, we evaluate on MS~COCO$_{\text{t2i}}$ and the video retrieval benchmark MSR-VTT. As shown in Fig.~\ref{fig:62_vision_as_document}, the two upper subfigures measure latency under increasing Top-$k$ with a single query, demonstrating that vision-first reuse generalizes effectively across both vision-to-text and text-to-vision reranking settings.
The lower subfigures evaluate latency under increasing numbers of queries with fixed Top-100 reranking, showing that the latency advantage of \emph{miniReranker} further amplifies as the reranking workload scales, reaching $<1\%$ of the original runtime on the long-video benchmark MSR-VTT.

\begin{figure}[h]
    \centering
    \includegraphics[width=1\linewidth]{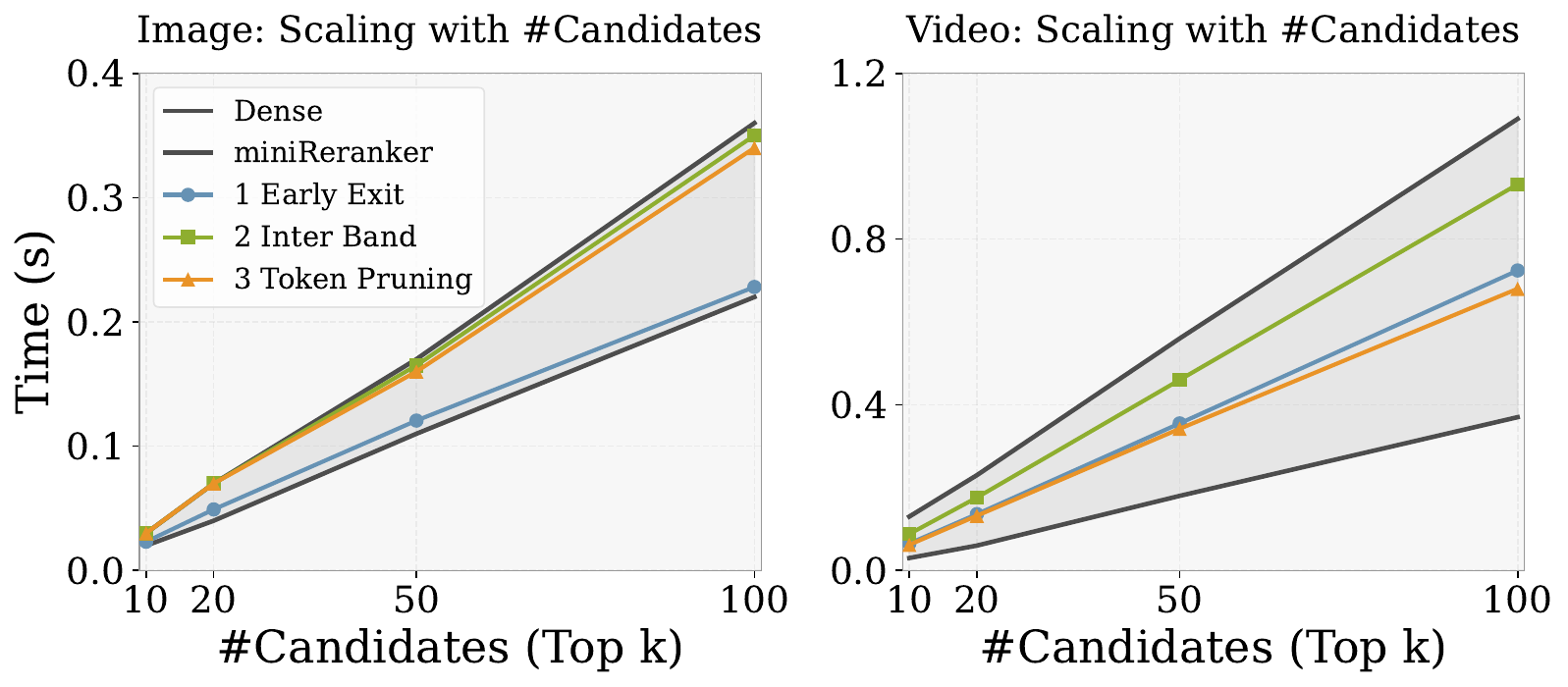}
    \vspace{-20pt}
    \caption{\small Latency breakdown of compression components.}
    \vspace{-10pt}
    \label{fig:63_ablation_vision}
\end{figure}

\paragraph{Reranking Latency: Ablation.}
We further analyze the contribution of each compression component to reranking acceleration. We scale the number of candidates and report the latency averaged over the two reuse scenarios and tasks. As shown in Fig.~\ref{fig:63_ablation_vision}, early exit provides the largest gain on image tasks, where the sequence length is relatively short and the cost is mainly dominated by model depth. In contrast, for video tasks with much longer visual sequences, visual token pruning and interaction-band restriction become more important, as they directly reduce the sequence length and cross-segment attention cost. Combining all components yields the lowest latency across both modalities, indicating that the proposed techniques address complementary sources of computation.

\paragraph{Overall,} \emph{miniReranker} consistently improves efficiency across training and inference. It reduces the active parameters to about $58\%$ of the dense model, achieves nearly $3\times$ training speedup, and substantially lowers reranking latency under both vision-as-query and vision-as-document settings. The reduction is especially pronounced for video reranking; in our measured settings, \emph{miniReranker} reduces the runtime to less than $1\%$ of the dense baseline. These results show that vision-first reuse, early exit, interaction-band restriction, and visual token pruning jointly provide an effective compression framework for efficient multimodal reranking.
\section{Conclusions}

In this work, we present \texttt{miniReranker}, an efficient MLLM-based point-wise reranking framework. By combining vision-first prompting with model-level compression, our method substantially reduces reranking computation while preserving most dense-model performance. Experiments across image, video, and visual-document tasks demonstrate strong efficiency--effectiveness trade-offs for scalable multimodal reranking.

\section*{Limitations}

Our work primarily focuses on point-wise multimodal reranking, where relevance is independently estimated for each query--document pair via prefill-only yes/no scoring. While this formulation is widely adopted in existing multimodal reranking systems, it does not capture richer interactions across candidates that may arise in list-wise reranking or agentic retrieval pipelines. Extending the proposed compression strategies to such settings remains an important direction for future work.

\section*{Ethical Considerations}

This work focuses on improving the efficiency of multimodal reranking systems through model compression and computation reuse, which may help reduce computational cost and energy consumption for large-scale deployment. Our method does not involve the collection of new user data or introduce additional ethical concerns beyond those already associated with the underlying multimodal foundation models and retrieval systems.



\bibliography{custom}

\clearpage

\appendix






\appendix



\section{Related Work}

\paragraph{MLLM-based Reranking and Retrieval.}
Recent multimodal retrieval systems increasingly adopt unified embedding models that support diverse modalities and instruction-conditioned retrieval~\cite{gme_GeneralMultimodalEmbedder,lin2025mmembed,qwen3-vl-embed-rerank}. 
While recent work explores MLLMs as rerankers using cross-encoder formulations. Prior studies show that instruction-tuned MLLMs can serve as effective multimodal relevance judges through point-wise \texttt{yes}/\texttt{no} scoring~\cite{vlm-as-reranker,liu2025lamra,lin2025mmembed,qwen3-vl-embed-rerank}. Other work additionally studies reasoning-enhanced reranking and application-specific multimodal retrieval settings~\cite{mm-r5,lu2026globalsimilarityfinegrainedmulticondition}. These studies collectively show the effectiveness of MLLM-based reranking. 
In contrast, our work targets the computational redundancy of point-wise multimodal reranking itself.

\paragraph{Efficient LLM-based Reranking.}
Prior work on efficient reranking mainly focuses on reducing autoregressive decoding, compressing long-context interactions, or caching document-side computation. Some methods replace generation with attention-based relevance estimation~\cite{chen2025attention,QRHead}, while others improve list-wise or long-context reranking through sparse attention, hierarchical ranking, or token pruning~\cite{zhou2026longranker,blockrank,sun2026efficientlistwisemultimodalreranking}. Another line of work reduces online reranking cost via document-side caching or compressed document representations~\cite{hyperRAG,dejean2025rerankingcompresseddocumentrepresentation,li2026efficientlongdocumentrerankingblocklevel}. While they mainly target text or document-centric reranking, our work provides a unified efficiency framework for multimodal point-wise rerankers.


\section{Early Exit: Layer-wise Logit Probing}
\label{appsec:early_exit}

\subsection{General VQA Tasks}
\label{appsubsec:early_exit_vqa}

To examine whether the early-exit behavior is specific to point-wise reranking, we further evaluate layer-wise probing on general VQA-style tasks

\begin{figure}[h]
    \centering
    \includegraphics[width=1.0\linewidth]{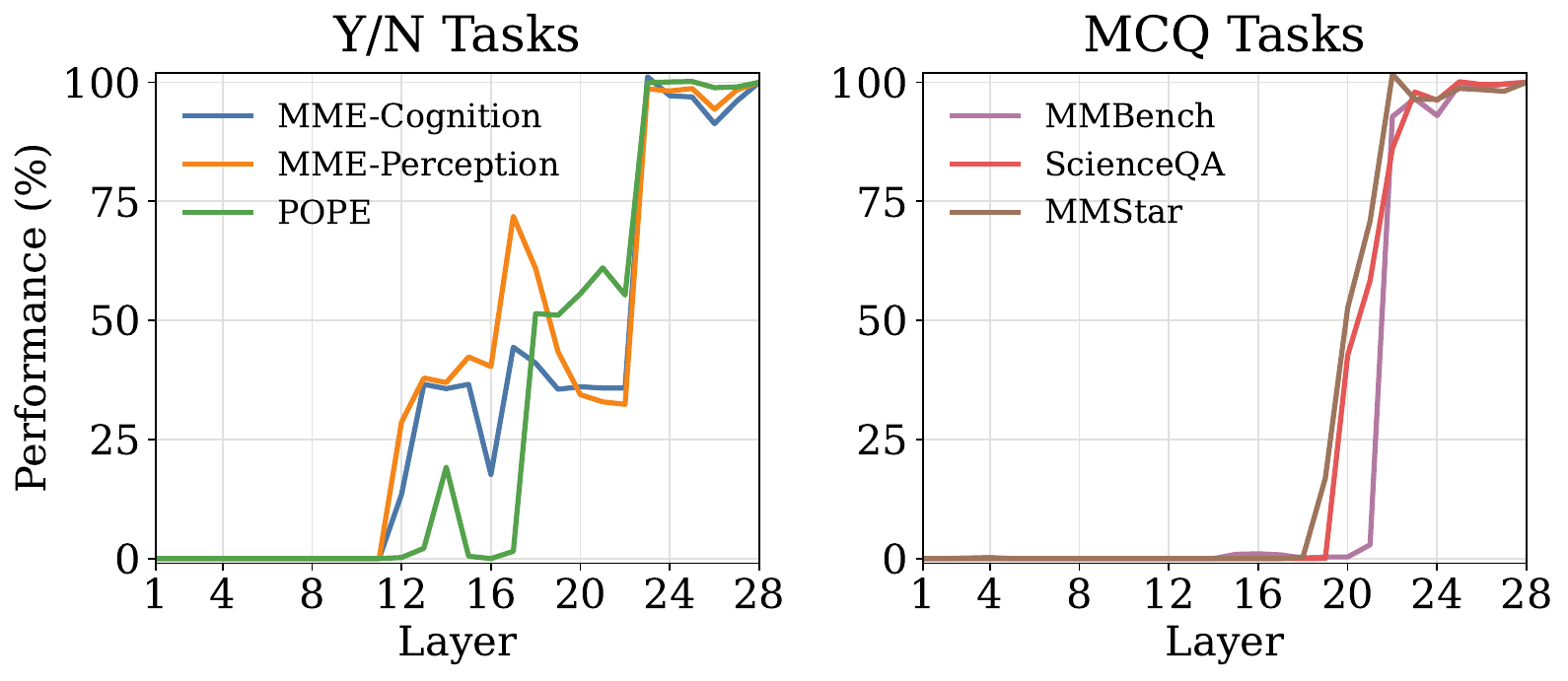}
    \vspace{-20pt}
    \caption{\small \textbf{Layer-wise probing on general VQA tasks.} We evaluate prefill-only yes/no tasks and multiple-choice tasks using intermediate-layer logits. Unlike point-wise reranking, general VQA tasks only recover final-layer performance at much deeper layers, typically around layer 22 or later.}
    \label{appfig:A2_general_vqa_early_exit_relative}
    \vspace{-15pt}
\end{figure}

\paragraph{Prefill-only Tasks.}
We consider two types of \emph{prefill-only} tasks: (1) yes/no tasks, including POPE~\cite{benchmark:pope} and MME~\cite{benchmark:mme}; and (2) multiple-choice tasks, including MMBench~\cite{benchmark:mmbench}, ScienceQA~\cite{benchmark:scienceqa}, and MMStar~\cite{benchmark:mmstar}. As shown in Fig.~\ref{appfig:A2_general_vqa_early_exit_relative}, both task types exhibit a much later emergence of reliable prediction signals than reranking. Across these datasets, the intermediate layers before around layer 20 generally fail to recover the final-layer performance, while performance only becomes comparable to the final layer at around layer 22 or later. This trend suggests that general VQA tasks require deeper-layer computation to form stable answer predictions, in contrast to point-wise reranking where strong relevance signals already emerge in intermediate layers.

\paragraph{Open-ended Tasks.}
We also evaluate open-ended VQA tasks, where the model needs to generate free-form answers. Since full layer-wise probing with decoding is computationally expensive, we only test representative early-exit layers selected in our reranking experiments. Specifically, we evaluate GQA~\cite{benchmark:gqa} and TextVQA~\cite{benchmark:textvqa} at the 16-th layer, which is effective for our point-wise reranking setting. However, early exiting at this layer fails to produce correct answers on these open-ended tasks, resulting in 0.0 scores on both datasets. This further indicates that open-ended generation depends more heavily on late-layer computation, including answer formulation and token-level decoding refinement.

Overall, these results show a clear difference between general VQA and point-wise reranking. While general VQA tasks require deeper layers to support answer prediction or generation, point-wise reranking mainly relies on relevance discrimination signals that emerge much earlier. This supports our motivation for applying early exit specifically to multimodal reranking rather than treating it as a generic compression strategy for all multimodal tasks.

\subsection{Finetuned Models}

We further analyze the layer-wise reranking behavior of Qwen3-VL-2B-Instruct$^{\dagger}$ after supervised finetuning under different prompt formulations. Specifically, we perform layer-wise logit probing on models finetuned with the query-first, document-first, and visual-first prompting strategies.

\begin{figure}[h]
    \centering
    \includegraphics[width=0.8\linewidth]{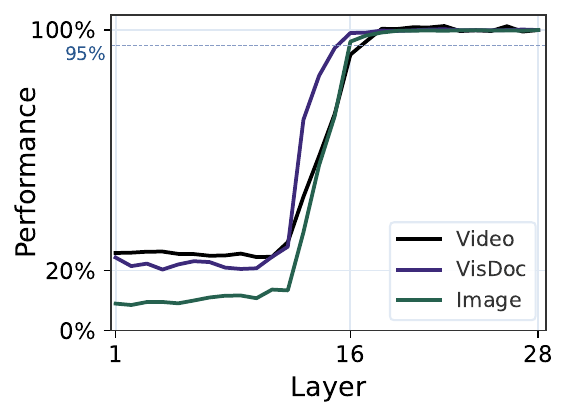}
    \vspace{-10pt}
    \caption{\small \textbf{Qwen3-VL-2B-reranker}$^{DF\dagger}$.}
    \label{appfig:A3_0_layer_probing_logits_DF}
    \vspace{-20pt}
\end{figure}

\begin{figure}[h]
    \centering
    \includegraphics[width=0.8\linewidth]{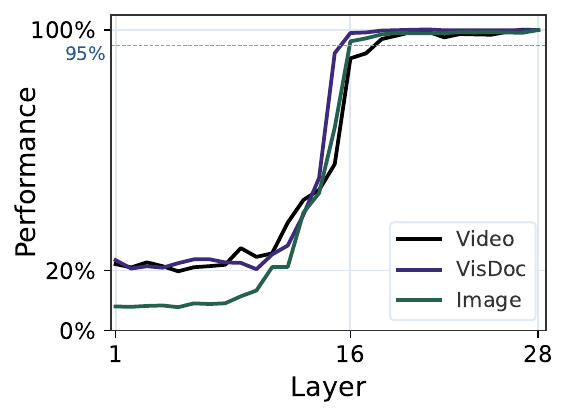}
    \vspace{-10pt}
    \caption{\small \textbf{Qwen3-VL-2B-reranker}$^{VF\dagger}$.}
    \label{appfig:A3_1_layer_probing_logits_VF}
    \vspace{-20pt}
\end{figure}

\begin{figure}[h]
    \centering
    \includegraphics[width=0.8\linewidth]{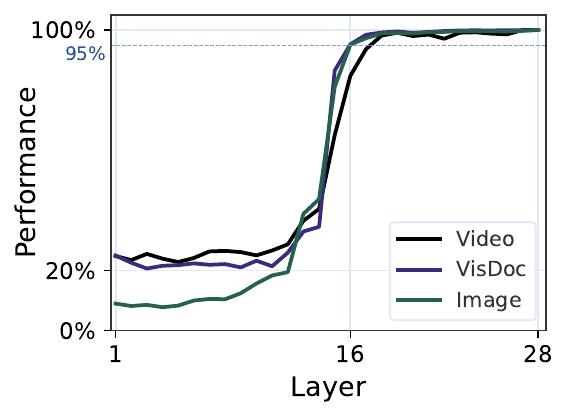}
    \vspace{-10pt}
    \caption{\small \textbf{Qwen3-VL-2B-reranker}$^{QF\dagger}$.}
    \label{appfig:A3_2_layer_probing_logits_QF}
    \vspace{-10pt}
\end{figure}

As shown in Fig.~\ref{appfig:A3_0_layer_probing_logits_DF}, Fig.~\ref{appfig:A3_1_layer_probing_logits_VF} and Fig.~\ref{appfig:A3_2_layer_probing_logits_QF} , finetuning with the binary \texttt{yes}/\texttt{no} reranking objective substantially changes the distribution of reranking signals across layers. Compared with the pretrained model, the finetuned models achieve stronger reranking performance at intermediate layers, reaching better performance around layer 16 while also exhibiting more stable behavior in deeper layers. This observation further supports our early-exit analysis, suggesting that reranking-relevant representations emerge relatively early in the transformer stack after task-specific finetuning. 

Moreover, the three prompt formulations exhibit highly similar layer-wise trends after finetuning. This suggests that prompt formatting mainly affects efficiency and computation reuse, while having relatively limited influence on where reranking signals emerge across transformer layers.



\section{Training Configuration}
\label{appsec:train}

\subsection{Token-Pruned Multimodal Reranker Training Set}

We construct a token-pruned multimodal reranker training set for learning fine-grained query-document relevance across heterogeneous modalities. Each example is represented as a query-document pair $(q, d)$ with a binary relevance label. Positive candidates are annotated with \texttt{label=yes}, while negative candidates are annotated with \texttt{label=no}. The negative set contains both hard negatives and randomly sampled negatives, enabling the reranker to learn both subtle semantic distinctions and broad modality-level discrimination.

\paragraph{Dataset Composition.}
Our training data is constructed from four multimodal sources: MMEB-train~\cite{benchmark:MMEB-v1} for image-related tasks, ViDoRe and VisRAG~\cite{dataset:vidore,dataset:visrag} for visual document retrieval, and ShareGPTVideo~\cite{dataset:sharegptvideo} for video-language retrieval. The resulting dataset contains 736,327 training pairs from 25 source subsets, covering a diverse collection of retrieval and reranking scenarios, including image-to-text, text-to-image, image-to-image, text-to-visual-document, text-to-video, video-to-text, and visual-question-to-text matching. In total, the dataset contains 147,899 positive pairs and 588,428 negative pairs, where the negative pairs consist of 294,375 hard negatives and 294,053 random negatives.

\begin{table}[t]
\centering

\footnotesize
\begin{tabular}{llr}
\toprule
Subset & Task Type & \#Samples \\
\midrule
A-OKVQA & Image+Text $\to$ Text & 8,525 \\
CIRR & Image+Text $\to$ Image & 13,055 \\
ChartQA & Image+Text $\to$ Text & 14,145 \\
DocVQA & Image+Text $\to$ Text & 19,730 \\
HatefulMemes & Image $\to$ Text & 1,700 \\
ImageNet-1K & Image $\to$ Text & 50,000 \\
InfographicsVQA & Image+Text $\to$ Text & 11,970 \\
MSCOCO & Image $\to$ Image & 50,000 \\
MSCOCO i2t & Image $\to$ Text & 56,640 \\
MSCOCO t2i & Text $\to$ Image & 50,000 \\
N24News & Image $\to$ Text & 24,490 \\
NIGHTS & Image+Text $\to$ Image & 7,970 \\
OK-VQA & Image+Text $\to$ Text & 4,500 \\
SUN397 & Image $\to$ Text & 9,925 \\
VOC2007 & Image $\to$ Text & 3,920 \\
VisDial & Text $\to$ Image+Text & 61,640 \\
Visual7W & Image+Text $\to$ Text & 34,905 \\
VisualNews i2t & Image $\to$ Text & 50,000 \\
VisualNews t2i & Text $\to$ Image+Text & 49,950 \\
WebQA & Text $\to$ Image+Text & 8,580 \\
\midrule
ViDoRe & Text $\to$ VisDoc & 59,095 \\
VisRAG & Text $\to$ VisDoc & 61,375 \\
\midrule
ShareGPTVideo t2v & Text $\to$ Video & 29,971 \\
ShareGPTVideo v2t & Video $\to$ Text & 29,998 \\
ShareGPTVideo QA & Video+Text $\to$ Text & 24,243 \\
\bottomrule
\end{tabular}
\vspace{-5pt}
\caption{Composition of the token-pruned multimodal reranker training set.}
\vspace{-20pt}
\label{tab:reranker_dataset_stats}
\end{table}

\paragraph{Negative Sampling.}
We convert each query into multiple point-wise training pairs by pairing every positive instance with four negatives, including two random negatives and two hard negatives. For image and visdoc tasks, the hard negatives are sampled from the top-20 retrieved candidates using the \texttt{Qwen3-VL-Embedding-2B}. For video tasks, the hard negatives are sampled from the top-5 retrieved candidates using the \texttt{Qwen3-VL-Embedding-2B}.

\paragraph{Data Schema.} Each sample follows a unified schema. The fields \texttt{subset} and \texttt{source\_row\_idx} identify the original data source and row index. The field \texttt{candidate\_type} specifies whether the candidate is a positive example, a hard negative, or a random negative. The fields \texttt{instruction}, \texttt{query\_text}, and \texttt{document\_text} provide the task instruction and textual content on the query and document sides. Visual inputs are stored either as image bytes, through \texttt{query\_image\_bytes} and \texttt{document\_image\_bytes}, or as sampled video-frame paths, through \texttt{query\_video\_frame\_paths} and \texttt{document\_video\_frame\_paths}. The field \texttt{doc\_modality} records the document-side modality, such as \texttt{text}, \texttt{image}, \texttt{image+text}, or \texttt{video}.

\paragraph{Token-Pruning Metadata.}
In addition to the original multimodal content, the dataset includes token-pruning metadata. The fields \texttt{query\_visual\_modalities} and \texttt{document\_visual\_modalities} indicate which visual modalities are present on each side. The fields \texttt{query\_visual\_token\_counts} and \texttt{document\_visual\_token\_counts} record the number of visual tokens for each visual input, while \texttt{query\_visual\_ranked\_tokens} and \texttt{document\_visual\_ranked\_tokens} store token-importance rankings. The token rankings are computed following~\autoref{subsec:compression}, by summing the attention scores from the last token to all visual tokens across all layers of \texttt{Qwen3-VL-2B-Embedding}. The rankings for visual inputs are saved during hard negative mining and therefore require no additional computation. These rankings allow the reranker to perform efficient visual token selection while preserving the original supervision signal of each query-document pair.

\subsection{Detailed Configuration}
\label{appsubsec:train}

We use \texttt{Qwen3-VL-Instruct} as the backbone model and fine-tune only the LLM component with a supervised point-wise reranking objective. All models are trained for one epoch using DeepSpeed ZeRO-2 with bfloat16 precision. The 2B model is trained on 4$\times$NVIDIA A100-SXM4-40GB GPUs, while the 4B and 8B models are trained on 4$\times$NVIDIA H100 GPUs. We use a per-device batch size of 4 with gradient accumulation steps of 1.

Optimization is performed with AdamW using a learning rate of $1\times10^{-4}$, zero weight decay, cosine learning rate scheduling, and a warmup ratio of 0.03. For parameter-efficient fine-tuning, we adopt LoRA with rank $r=16$, scaling factor $\alpha=32$, and dropout rate 0.05. Unless otherwise specified, all experiments use the same training configuration.

\section{Evaluation}
\label{appsec:eval}

\subsection{MMEB-V2}
\label{appsec:tasks}

We provide a detailed summary of the evaluation tasks in Tab.~\ref{apptab:all_tasks}. The benchmark covers three major multimodal domains: image-centric tasks, video-centric tasks, and visual document retrieval. For each task group, we report the meta-task category, source datasets, retrieval direction, number of queries, and candidate pool size. The retrieval direction is denoted by the input and target modalities, where I, T, V, and D represent image, text, video, and visual document, respectively. 

Overall, the evaluation includes a broad range of multimodal retrieval and reranking scenarios, such as image classification, visual question answering, image retrieval, visual grounding, video retrieval, moment retrieval, video classification, video question answering, and visual document retrieval. This diverse task coverage allows us to evaluate whether the proposed reranking framework generalizes across different modality combinations, retrieval directions, and candidate set sizes.

\begin{table}[t]
\centering
\scriptsize
\setlength{\tabcolsep}{3pt}
\renewcommand{\arraystretch}{1.4}
\vspace{-10pt}
\resizebox{\columnwidth}{!}{
\begin{tabular}{p{0.18\linewidth} p{0.43\linewidth} p{0.18\linewidth} p{0.10\linewidth} p{0.11\linewidth}}
\toprule
\textbf{Meta-Task} & \textbf{Dataset} & \textbf{Retrieval} & \textbf{\#Query} & \textbf{\#Cand.} \\
\midrule
Image Classification 
& ImageNet-1K, N24News, HatefulMemes, VOC2007, SUN397, Place365, ImageNet-A/R, ObjectNet, Country-211
& I$\to$T, I+T$\to$T
& 1,000
& 2--1,000 \\
Image VQA
& OK-VQA, A-OKVQA, DocVQA, InfographicVQA, ChartQA, Visual7W, ScienceQA, VizWiz, GQA, TextVQA
& I+T$\to$T
& 1,000
& 1,000 \\

Image Retrieval
& VisDial, CIRR, VisualNews, MSCOCO, NIGHTS, WebQA, OVEN, FashionIQ, EDIS, Wiki-SS-NQ
& T$\to$I, I$\to$T, I+T$\to$I/T
& 1,000
& 1,000 \\

Visual Grounding
& MSCOCO, Visual7W-Pointing, RefCOCO, RefCOCO-Matching
& I+T$\to$I/T
& 1,000
& 1,000 \\
\midrule
Video Retrieval
& DiDeMo, MSR-VTT, MSVD, VATEX, YouCook2
& T$\to$V
& 670--4,468
& 670--4,468 \\

Moment Retrieval
& QVHighlights, Charades-STA, MomentSeeker
& T+V$\to$V
& 727--1,800
& 10 \\

Video Classification
& Kinetics-700, SSv2, HMDB51, UCF101, Breakfast
& V$\to$T
& 433--1,000
& 10--700 \\

Video QA
& MVBench, Video-MME, NExT-QA, EgoSchema, ActivityNetQA
& V+T$\to$T
& 500--8,564
& 2--5 \\
\midrule
Visual Document Retrieval
& ViDoRe, ViDoRe-V2, VisRAG, ViDoSeek, MMLongBench-Doc
& T$\to$D
& 52--1,646
& 70--9,590 \\
\bottomrule
\end{tabular}
}
\vspace{-8pt}
\caption{Evaluation dataset composition. We summarize the evaluated datasets by meta-task, retrieval format, number of queries, and candidate size. I, T, V, and D denote image, text, video, and visual document, respectively.}
\label{apptab:all_tasks}
\vspace{-20pt}
\end{table}

\subsection{Baselines}
\label{appsubsec:baselines}

\paragraph{ColPali v1.3}
ColPali is a specialized visual document retriever designed for page-level document retrieval. Instead of relying on OCR-based text extraction, it directly embeds document page images with a VLM and produces ColBERT-style multi-vector representations for late-interaction matching~\cite{dataset:vidore}. In our comparison, ColPali v1.3 serves as a strong visual-document retrieval baseline, particularly for tasks where layout, tables, figures, and page-level visual cues are important.

\paragraph{General Multimodal Embedder (GME)}
GME is an instruction-aware multimodal dense retriever built on MLLMs. It represents text, images, visual documents, and image-text composed inputs in a unified embedding space, and is trained with contrastive learning over diverse single-modal, cross-modal, and fused-modal retrieval data~\cite{gme_GeneralMultimodalEmbedder}. We include GME as a general-purpose multimodal embedding baseline, since it supports a broad range of retrieval scenarios beyond visual document retrieval.

\paragraph{VLM2Vec}
VLM2Vec is a contrastive training framework that converts existing vision-language models into universal multimodal embedding models~\cite{benchmark:MMEB-v2}. Unlike CLIP-style models that encode images and text independently without task instructions, VLM2Vec can process arbitrary combinations of image and text inputs and produce task-conditioned fixed-dimensional embeddings. We use it as an instruction-aware multimodal embedding baseline trained for diverse embedding tasks such as classification, VQA, multimodal retrieval, and visual grounding.

\paragraph{Qwen3-VL-Reranker-2B}
Qwen3-VL-Reranker-2B is a multimodal cross-encoder reranker built on Qwen3-VL. Given a query-document pair, where both sides may contain text, images, screenshots, videos, or mixed-modal inputs, it performs fine-grained relevance estimation and outputs a relevance score~\cite{qwen3-vl-embed-rerank}. Compared with embedding-based baselines, this model is closer to our setting because it jointly encodes each query-document pair for reranking. We therefore use it as the dense reranker baseline before applying our compression strategy.

\subsection{More detailed configurations}
Following Qwen3-VL-Reranker~\cite{qwen3-vl-embed-rerank}, we constrain image inputs to 4–1800 visual tokens (4096–1,843,200 pixels), and video inputs to at most 64 frames sampled at 1 FPS with a total budget of 7,864,320 pixels; the maximum input length is capped at 10,240 tokens.

During evaluation, we use the raw logit difference between the \texttt{yes} and \texttt{no} tokens as the reranking score, instead of applying a sigmoid normalization. Since the sigmoid function is monotonic, this does not change the ranking order. In practice, using raw logits avoids numerical saturation when the logit difference is very large or very small, which can otherwise make multiple samples receive indistinguishable scores after sigmoid normalization. To further ensure evaluation robustness, we randomly shuffle the candidate order before reranking, so that the results are not affected by the retrieval model's original candidate ordering.

\subsection{Prompt Templates}
\label{appsec:prompt_templates}

We use a unified point-wise prompt format for all reranking tasks, where the model is required to judge whether a candidate document satisfies the given query and instruction by generating only \texttt{yes} or \texttt{no}. Since the reusable visual segment differs across retrieval settings, we adopt two visual-first variants depending on whether the visual input is associated with the query or the document.

\paragraph{Vision-as-query.}
For tasks where the visual input belongs to the query, e.g., image-to-text or video-to-text retrieval, we place the query before the candidate document. This allows the query-side visual representations to be computed once and reused across all candidate documents for the same query.

\begin{figure}[h]
\centering
\begin{promptbox}{Input Template for Vision-as-query Reranking}
<|im_start|>system
Judge whether the Document meets the requirements based on the Query and the Instruct provided.
Note that the answer can only be "yes" or "no".
<|im_end|>

<|im_start|>user
<Instruct>: {Instruction}
<Query>: {Query}
<Document>: {Document}
<|im_end|>
<|im_start|>assistant
\end{promptbox}
\vspace{-6pt}
\caption{\small Prompt template for vision-as-query reranking, where the query-side visual input is placed before the candidate document to enable reuse across candidates.}
\label{appfig:prompt_template_vq}
\end{figure}

\paragraph{Vision-as-document.}
For tasks where the visual input belongs to the document, e.g., text-to-image or image-to-image retrieval, we instead place the document before the query. This makes the document-side visual representations independent of the incoming query, enabling reuse across different queries.

\begin{figure}[h]
\centering
\begin{promptbox}{Input Template for Vision-as-document Reranking}
<|im_start|>system
Judge whether the Document meets the requirements based on the Query and the Instruct provided.
Note that the answer can only be "yes" or "no".
<|im_end|>

<|im_start|>user
<Instruct>: {Instruction}
<Document>: {Document}
<Query>: {Query}
<|im_end|>
<|im_start|>assistant
\end{promptbox}
\vspace{-6pt}
\caption{\small Prompt template for vision-as-document reranking, where the document-side visual input is placed before the query to enable reuse across queries.}
\label{appfig:prompt_template_vd}
\end{figure}







\section{Ablation on Visual Token Selection}
\label{appsec:selector}

To further validate the effectiveness of our embedder-attention-guided token selection strategy, we compare it with several alternative visual token selection methods. For fair comparison, all methods prune visual tokens before they are fed into the reranker LLM, and therefore reduce the same subsequent LLM-side computation. We only consider image-intrinsic selection criteria that depend on the visual input itself, rather than query-document relevance labels or external supervision. In addition, none of the compared methods introduces extra model parameters or additional forward passes. Under this setting, we evaluate two representative alternatives: similarity-based selection using the embedder representations and attention-based selection using the reranker's vision encoder.

\subsection{\texttt{<Emb>} Similarities}

\paragraph{Methods.}
A natural alternative is to select visual tokens according to their similarity to the final embedding token. Specifically, we use the hidden representation of the final \texttt{<Emb>} token produced by the retrieval-stage embedder as a global summary of the input, and compute its cosine similarity with each visual token representation. Visual tokens with higher similarity scores are retained. We evaluate this strategy using representations from different layers, including the 3rd layer, the 14th layer, the last layer, and an all-layer variant that aggregates similarity scores across layers.

\paragraph{Results.}
As shown in Tab.~\ref{apptab:embedder_simi_selection}, similarity-based selection provides a simple way to estimate visual token importance, but it is less effective than our attention-based embedder-guided strategy. This suggests that direct representation similarity to the final \texttt{<Emb>} token does not fully capture which visual tokens are most useful for reranking. In contrast, our method uses the attention distribution from the final embedding token, which better reflects how the embedder aggregates visual evidence for retrieval-oriented representation learning.

\begin{table}[h]
\footnotesize
\centering
\vspace{-8pt}
\renewcommand{\arraystretch}{1.2}
\resizebox{0.48\textwidth}{!}{
\begin{tabular}{lccccc}
\toprule
& \multicolumn{5}{c}{Image} \\
\cmidrule(lr){2-6}
Selection & CLS & QA & RET & GD & \textbf{Overall} \\
\midrule
\rowcolor{gray!15} Dense & 59.1 & 61.2 & 51.4 & 77.3 & 59.1 \\
Simi(3rd Layer)  & 56.7 & 54.6 & 50.2 & 65.8 & 54.9 \\
Simi(14th Layer) & 56.9 & 58.3 & 50.2 & 65.4 & 56.0 \\
Simi(Last Layer) & \textbf{58.3} & 58.8 & 50.5 & 63.5 & 56.4 \\
Simi(All Layers) & 57.6 & 58.9 & 50.3 & 64.5 & 56.3 \\
\rowcolor{TP!15} \textbf{Ours} & 58.1 & \textbf{60.8} & \textbf{51.2} & \textbf{72.6} & \textbf{58.2} \\
\bottomrule
\end{tabular}
}
\vspace{-8pt}
\caption{Ablation of similarity-based visual token selection using the embedder \texttt{<Emb>} representation.}
\label{apptab:embedder_simi_selection}
\vspace{-20pt}
\end{table}

\subsection{Reranker ViT Attention}

We also compare with token selection based on the reranker's own vision encoder. This method ranks visual tokens by their attention scores inside the ViT, where tokens receiving larger accumulated attention are regarded as more important. Similar to the similarity-based setting, we evaluate attention scores from the 3rd layer, the 14th layer, the last layer, and an all-layer aggregation variant. Since these scores are obtained from the reranker's visual encoding process itself, this baseline does not require an additional model or extra forward computation.

\paragraph{Results.}
Tab.~\ref{apptab:vit_attn_selection} reports the comparison with ViT-attention-based selection. Although ViT attention captures visual saliency within the reranker's vision encoder, it is not explicitly optimized for retrieval or reranking. As a result, visually salient tokens are not always the most useful tokens for relevance estimation. Our embedder-guided strategy performs better because the selection signal comes from the retrieval-stage embedder, whose attention patterns are more directly aligned with the downstream reranking objective.

\begin{table}[h]
\footnotesize
\centering
\vspace{-8pt}
\renewcommand{\arraystretch}{1.2}
\resizebox{0.48\textwidth}{!}{
\begin{tabular}{lccccc}
\toprule
& \multicolumn{5}{c}{Image} \\
\cmidrule(lr){2-6}
Selection & CLS & QA & RET & GD & \textbf{Overall} \\
\midrule
\rowcolor{gray!15} Dense & 59.1 & 61.2 & 51.4 & 77.3 & 59.1 \\
ViT(3rd Layer)  & 54.7 & 57.5 & 48.7 & 62.3 & 54.3 \\
ViT(14th Layer) & 54.6 & 58.0 & 47.4 & 71.4 & 55.0 \\
ViT(Last Layer) & 57.4 & 57.1 & \textbf{51.2} & \textbf{73.4} & 57.0 \\
ViT(All Layers) & 56.3 & 59.0 & 48.9 & 69.3 & 56.0 \\
\rowcolor{TP!15} \textbf{Ours} & \textbf{58.1} & \textbf{60.8} & \textbf{51.2} & 72.6 & \textbf{58.2} \\
\bottomrule
\end{tabular}
}
\caption{Ablation of visual token selection using attention scores from the reranker's ViT.}
\vspace{-10pt}
\label{apptab:vit_attn_selection}
\end{table}

\section{Efficiency Analysis}

\subsection{Precache Time}

Main-text latency results focus on \emph{online reranking latency}, where visual representations are assumed to be precomputed and cached. This setting reflects practical retrieval systems, in which visual features are typically reused across many reranking pairs. Moreover, cache construction can be performed either offline or online depending on the deployment setting, making the preprocessing overhead system-dependent and amortizable across repeated reuse. For completeness, we further report end-to-end latency including visual pre-encoding and cache construction overhead. Although precaching introduces additional preprocessing cost, the overhead can be amortized through repeated reuse during reranking.

\begin{figure}[h]
    \centering
    \includegraphics[width=1\linewidth]{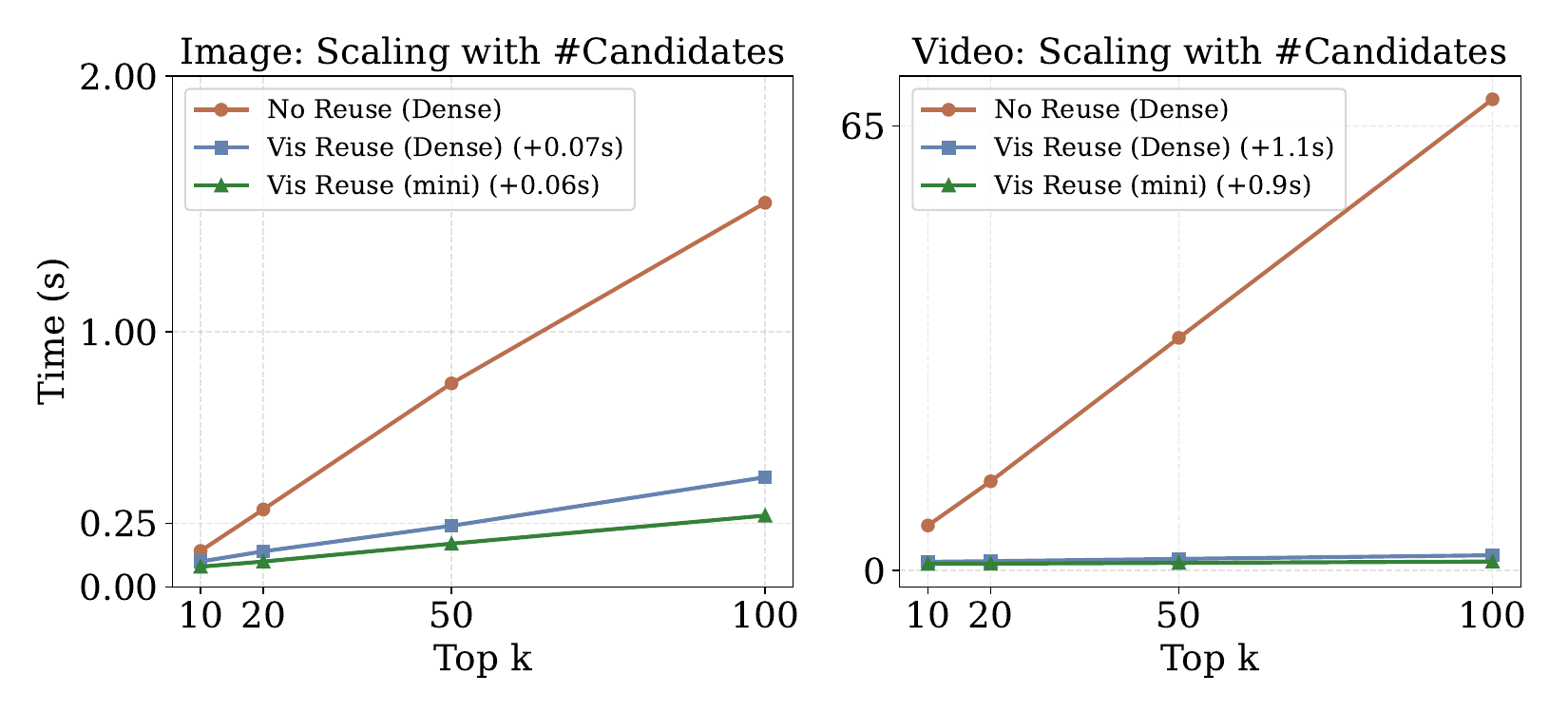}
    \vspace{-25pt}
    \caption{\small End-to-end latency including visual pre-encoding and cache construction overhead in the vision-as-query setting, measured under different numbers of candidate documents.}
    \vspace{-10pt}
    \label{appfig:d1_precache_v_query}
\end{figure}

\paragraph{Vision-as-query setting.} 
In this setting, the visual query representation is precached once and reused across multiple candidate documents. We therefore measure the end-to-end latency including both cache construction and reranking computation while varying the number of candidates. As shown in Fig.~\ref{appfig:d1_precache_v_query}, although precaching introduces additional preprocessing overhead, the cost remains relatively small compared with the overall reranking computation. The overhead is further amortized as the number of candidate documents increases, leading to substantially lower end-to-end latency than the vanilla reranking pipeline.

\begin{figure}[h]
    \centering
    \includegraphics[width=1\linewidth]{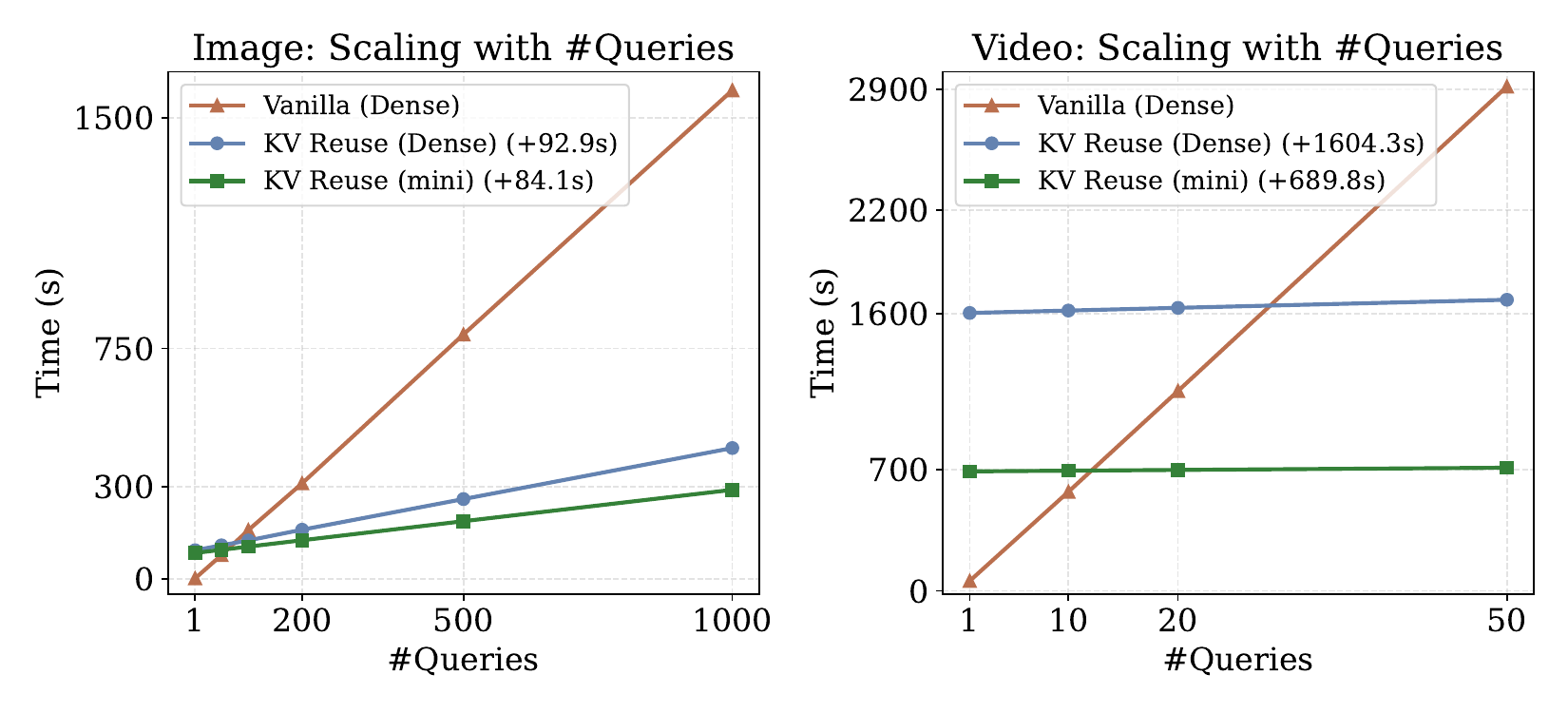}
    \vspace{-25pt}
    \caption{End-to-end latency including visual pre-encoding and cache construction overhead in the vision-as-document setting, measured under different numbers of queries.}
    \label{appfig:d2_precache_v_doc}
\end{figure}

\paragraph{Vision-as-document setting.} 
In this setting, document-side visual representations are cached once and reused across many incoming queries. Since the reuse frequency increases with the number of queries, we report latency scaling with respect to the number of queries. As shown in Fig.~\ref{appfig:d2_precache_v_doc}, the upfront cache construction cost becomes increasingly negligible as reuse grows. Consequently, despite including visual pre-encoding and cache construction overhead, the proposed reuse strategy still achieves substantially lower end-to-end latency compared with the vanilla reranking pipeline, particularly in large-scale retrieval scenarios.

\end{document}